# Superconductivity controlled by the magnetic state of ferromagnetic nanoparticles

A.A. Fraerman[1], B.A. Gribkov[1], S.A. Gusev[1], E. Il'ichev[2], A.Yu. Klimov[1], Yu.N. Nozdrin[1], G.L. Pakhomov[1], V.V. Rogov[1], R. Stolz[2] and S.N. Vdovichev[1]

[1] *Institute for Physics of Microstructures RAS, GSP 105, 603950 Nizhny Novgorod, Russia*

[2] *Institute for Physical High Technology, Jena, Germany*

Novel hybrid superconductor/ferromagnetic particles structures are presented. Arrays of Co submicron particles were fabricated on overlap, edge type Josephson junctions and on a narrow Nb microbridge by means of high-resolution e-beam lithography. We observed a strong dependence between the magnetic state of the particles and the field-dependent critical $I_c(H)$ of these structures.

## INTRODUCTION

Recent advances in microfabrication techniques have increased the interest in application of submicron sized patterned magnetic elements. The prospective usage of patterned ferromagnetic particles of submicron sizes is accounted for by the well defined local magnetic fields. In particular, magnetic dots arrays can be used as artificial pinning centers in a superconductor thin film. Experimental investigation of such a hybrid superconductor/ferromagnetic particles system was begun in the last decade from pioneer works of D. Givord et al. [1,2] and is underway currently, for example [3,4]. Transport properties of a superconducting layer may also assist in determining the magnetic states of a ferromagnetic particles array [2]. However, the magnetic mechanism of pinning in thin films works close to $T_c$, when "natural" pinning forces are insignificant [4,5]. Another way for controlling the superconducting state of a film by magnetic circuit is through modulating the superconducting order parameter and creating weak links [6]. This device can work as a superconductor switch.

In this paper, we present three novel hybrid structures: weak links - Josephson junctions (overlap and edge type) and a narrow superconductor microbridge patterned by magnetic dots arrays. The influence of the particle magnetization distribution on the static field-dependent critical current $I_c(H)$ of these structures has been investigated. All measurements were done at temperatures well below the superconducting transition.

## EDGE-TYPE JOSEPSON JUNCTION

*Theory.* We propose to use the influence of a stray magnetic field of nanoparticles on quantum interference in a Josephson junction. We model a real structure as an infinite strip of superconductor cut by a narrow slot, see Fig. 1. Assuming a sinusoidal current-phase relationship, the field-dependent total current $I(H)$ of the junction is expressed as

$$I(H) = \iint_S j_c e^{i\varphi(H)} dS ,$$

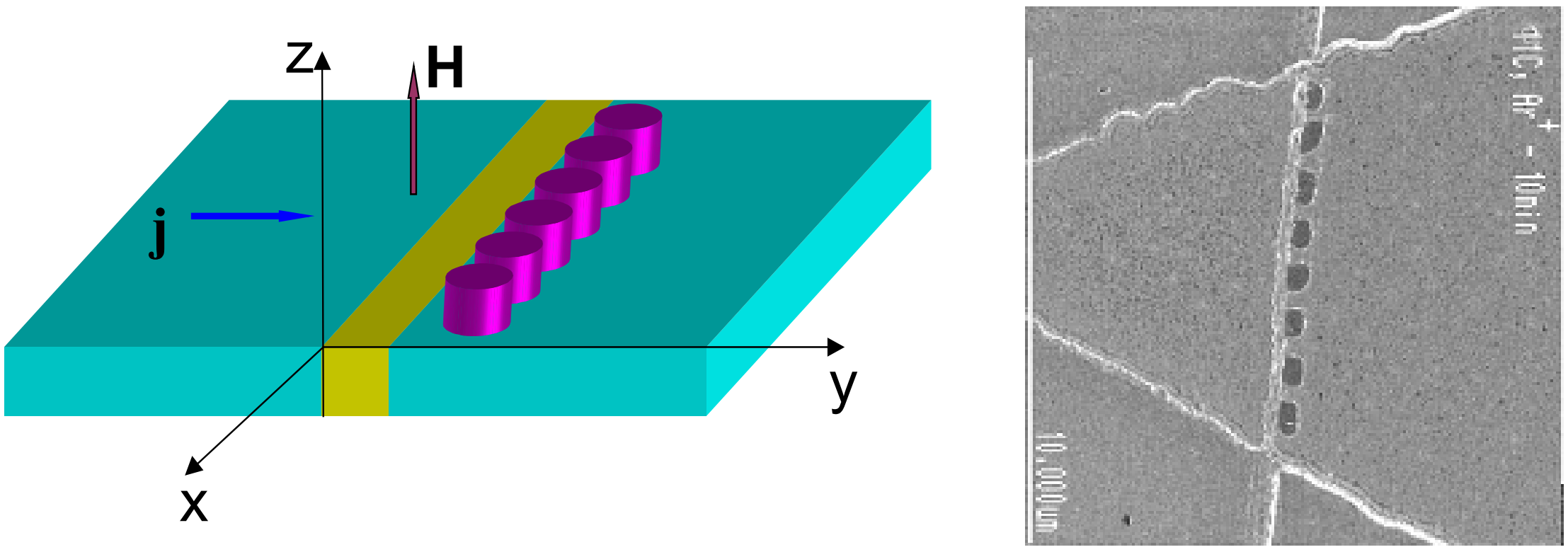

Fig. 1. Schematic and SEM image of an edge-type Josephson junction with a chain of magnetic nanoparticles.

where $\varphi$ is the phase differences, $j_c$ is the critical current density [7]. If the ferromagnetic particles are placed near the barrier, see Fig. 1, $\varphi$ can be written as

$$\varphi = \varphi_{ext} + \varphi_{particle},$$

where $\varphi_{ext}$, $\varphi_{particle}$ are the phase differences due to the external magnetic field and the magnetic field of the particles, respectively. It means that the magnetic field of the particles can change the Fraunhofer diffraction pattern $I_c(H)$. We now assume that the spatial distribution of the Josephson phase difference $\varphi(x)$ across the junction barrier can be written as

$$\frac{\partial \varphi}{\partial x} = \frac{2\pi\Lambda}{\Phi_0}\left(H + H_p\right),$$

where $\Lambda$ is the effective thickness of the junction ($\Lambda \approx 2\lambda_L$, $\lambda_L$ is the London penetration depth), $H$ is the external magnetic field, $H_p$ is the magnetic field induced by the magnetic particles, $\Phi_0 = hc/e = 2\cdot 10^{-7} Oe\cdot cm^2$ is the magnetic flux quantum.

In particular, if the magnetic moment of a single particle in the chain has a uniform distribution, $H_p(x)$, $\varphi_{particle}$ are periodic functions. The results of numeric simulation of $I_c(H)$ for a junction with five dipoles is shown on Fig. 2a. In this case the field-dependent critical current should have a maximum when the magnetic flux per period of the array of nanoparticles at the Josephson junction contains an integer number of flux quanta $\Phi_d = n\Phi_0$, ($\Phi_d = H\Lambda d$, $n$ is integer). A more explicit model for this case confirmed the same result [8].

*Experiment.* For experimental investigation of the dependence of the particles magnetization distribution on the critical current we fabricated a series of Nb\SiN$_x$\Nb edge-type Josephson junctions [9] with chains of ferromagnetic *Co* nanoparticles with a typical lateral size of 300-600 nm and a height of 25 nm (see Fig. 1). Measurements of the critical current $I_c(H)$ were performed by a standard four-terminal method at $T=4.2K$, in the magnetic field normal to the plane of the junction.

The particles were magnetized at room temperature either by applying a magnetic field of *20 kOe* or by the MFM tip [10]. The magnetic state of the nanoparticles was controlled by MFM before and after the low temperature experiments.

We present two series of experiments. 1). All particles were magnetized as dipoles (see inset on Fig. 2b) 2). All particles were magnetized close to the vortex state. In the case of the dipole magnetic states, we observed an increase in the additional $I_c(H)$ maxima whose position depends on a period of the particles chain, as theory predicted, see Fig. 2. In the case of the vortex magnetic states of the particles, the resonance effects were absent. This result is brought about by the weak magnetic field induced by the particles with a curling magnetization distribution.

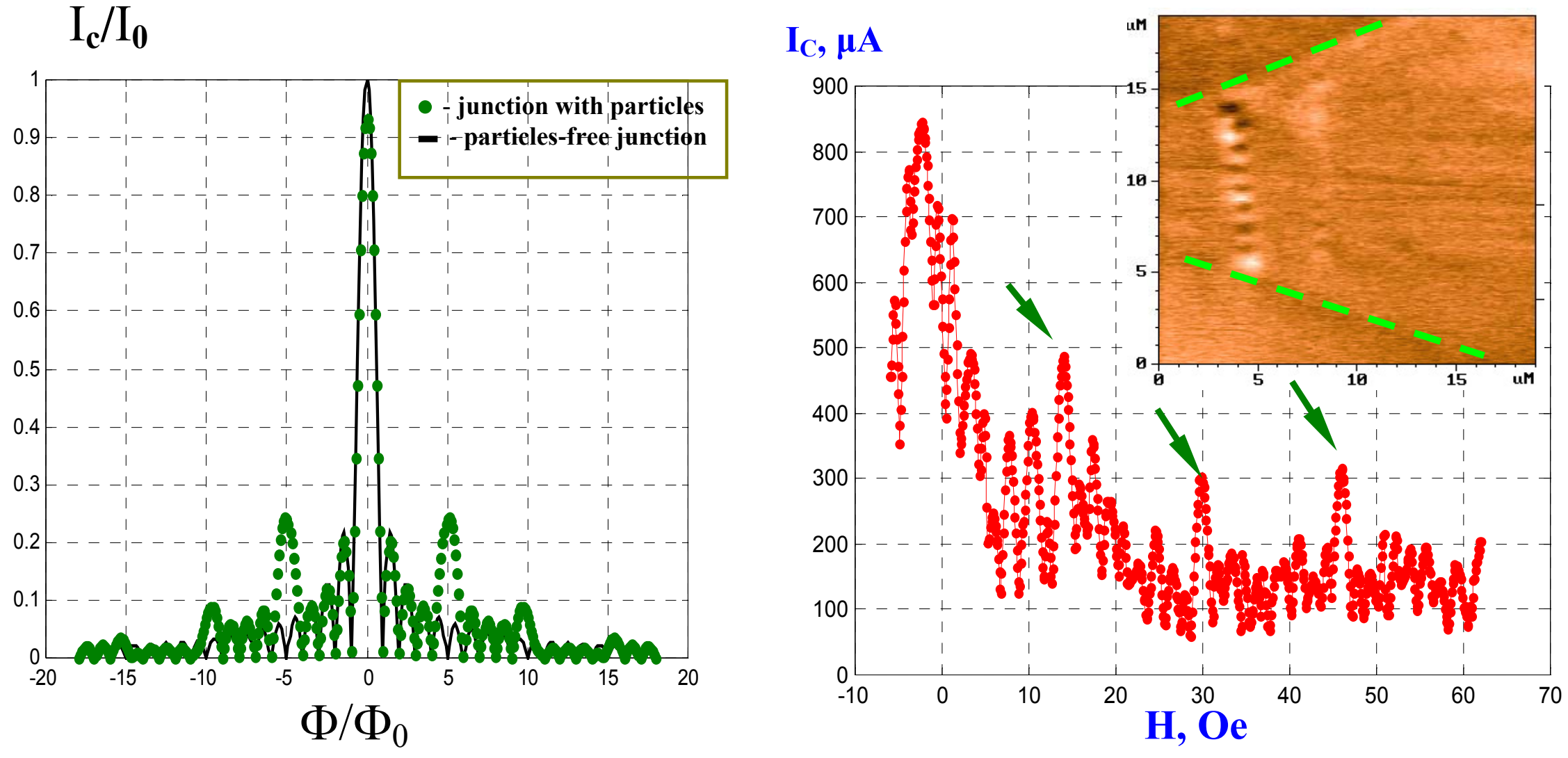

a) b)

Fig. 2. Dependence $I_c(H)$ of the junction with dipoles a) the numeric simulation b) the experiment (on the inset MFM image of the junction)

## OVERLAP JOSEPHSON JUNCTION

The basic idea of this experiment is to trap a regular lattice of the Abrikosov vortices in the top electrode of the overlap junction perpendicularly to the junction surface. This is a unique way for controlling the properties of the overlap type junction because neither the magnetic field induced by particles, nor the aligned vortices trapped in both electrodes affect the phase differences, i.e., properties of the junction. A special series of overlap geometry junctions *Nb\Al\AlO*$_x$*\Nb* with a thin top electrode (30 nm) was produced by IPHT (Jene, Germany). The width and length of the junctions is about *20 μm*. The top electrode thickness is smaller than $\lambda_L$ of the *Nb*, which makes easier the entrance of the Abrikosov vortices. The particles array was fabricated on the top of the junction, using electron beam lithography, see Fig. 3.

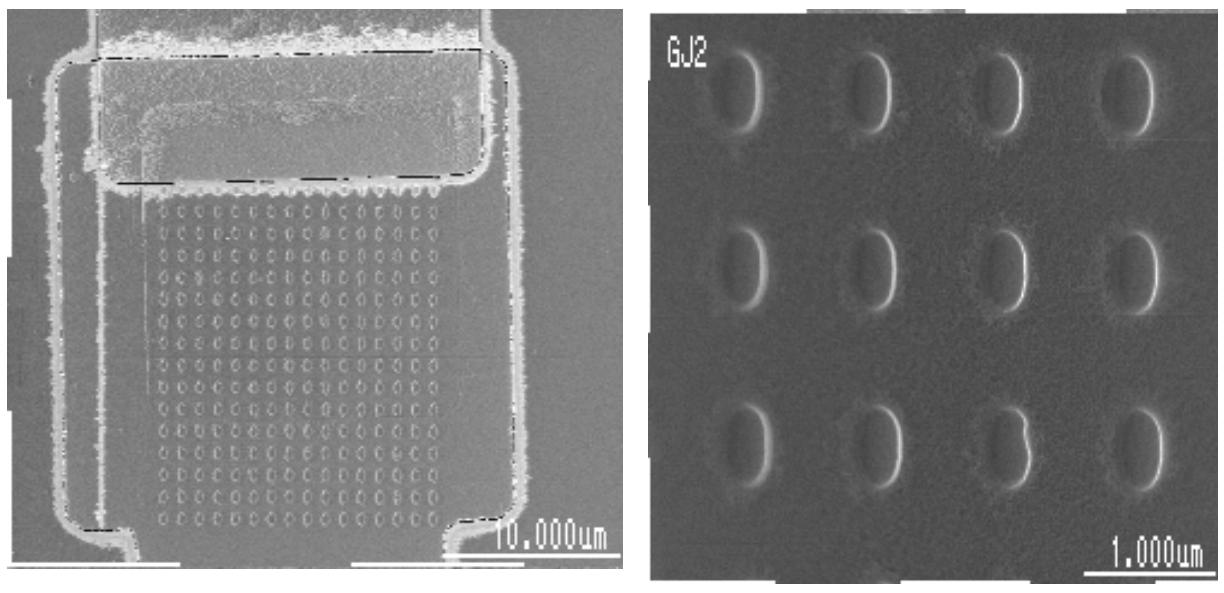

Fig. 3. SEM image of the overlap Josephson junction with an array of magnetic nanoparticles.

Measurements of the critical current $I_c(H)$ were performed by the four-terminal method for helium temperature, in the magnetic field within the plane of the junction.

It is evident that the trapping of vortices is controlled by the magnetic state of the particles. We also present two series of experiments. 1). About *80%* of the particles were magnetized as dipoles. 2). About *80%* of the particles were magnetized as vortices. The Fraunhofer pattern for the vortex magnetic state of the particles and for junction without particles is the same, see Fig. 4a. The diffraction pattern for the dipole magnetic state of the dots has changed considerably, see Fig. 4b. The critical current has been suppressed; the form of the Fraunhofer pattern has changed and become different from zero at a high magnetic field. There are no resonance effects that can be explained by the absence of the regular vortex lattice. The results observed in this experiment slightly resemble those obtained in a special experiment for a single vortex motion [11].

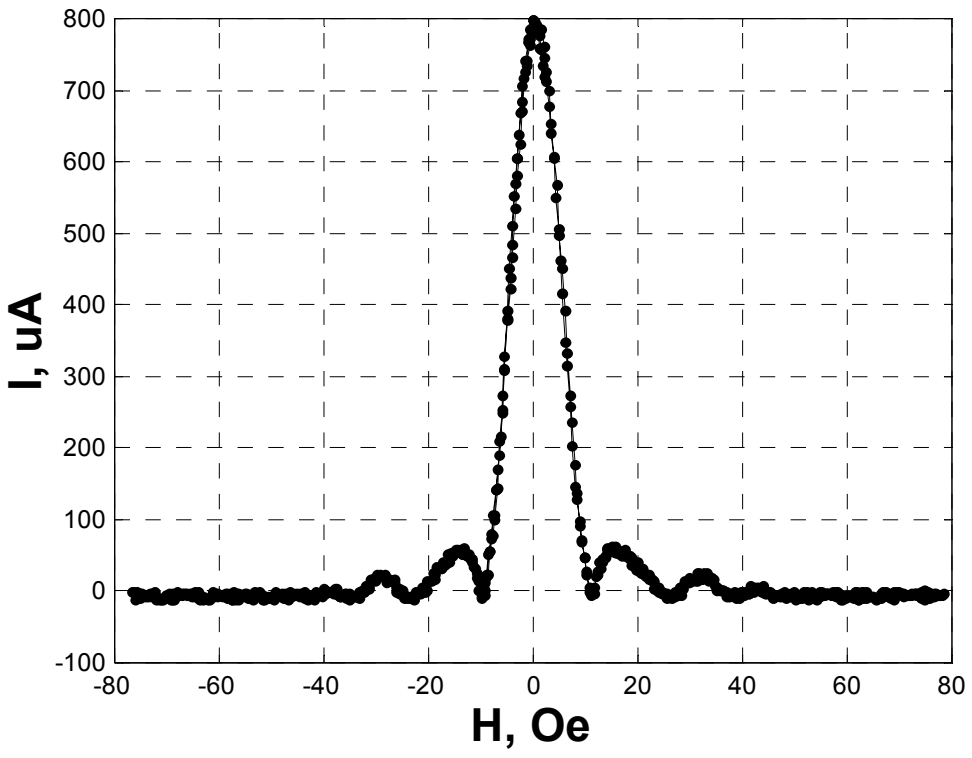

a)

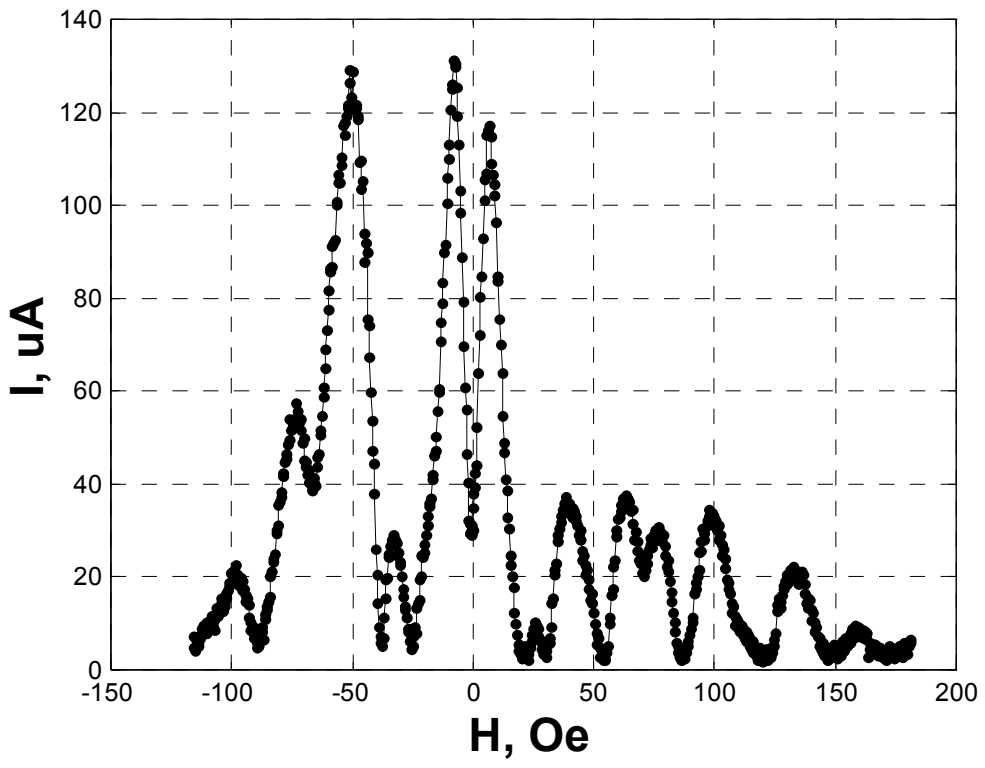

b)

Fig. 4. Dependence $I_c(H)$ of the overlap junction with a) prevalence of dipoles states, b) prevalence of vortex magnetic states.

## NARROW SUPERCONUCTOR MICROBRIDGE

We fabricated a series of *1 μm* wide *Nb* microbridges with the *Nb* thickness of about *0.1 μm* ($T_c$*~9 K*) and formed a chain of *Co* nanoparticles with a typical lateral size of *300*600 nm* and a height of *100 nm* (Fig. 5a). The remanent magnetic state of the particles is inhomogeneous; particles can be magnetized to a uniform state by the field of *500-1000 Gs* [13]. The density of the critical current for our *Nb* films is about $5\ 10^7\ A/cm^2$, which implies deep centers of pinning [14].

We performed standard measurements of the critical current $I_c(H)$ at $T=4.2K$ in an external magnetic field normal to the plane of the microbridge.

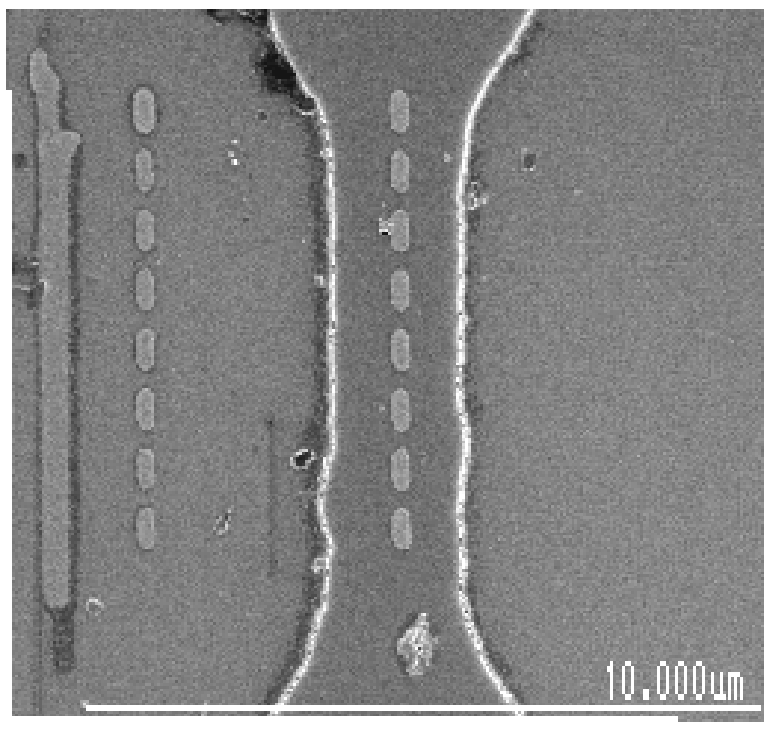

Fig. 5. SEM image of microbridges with ferromagnetic nanoparticles.

The critical current dramatically decreased with an increasing magnetic field and this dependence was the same for samples with and without particles. If a magnetic field is applied in the plane of a particle-free microbridge (up to *2.5 kOe*), the critical current remains unaffected. For a microbridge with magnetic particles we observed the following effects, see Fig. 6: enhancement of the critical current $I_c$ to *20%* with an increasing magnetic field *H*; strong dependence of the critical current on the magnetic field sign, the difference between $I_c(H)$ and $I_c(-H)$ reached *70%* ("diode effect"). In addition, the critical current has a periodical component with a period about *500 Oe*, see Fig. 6.

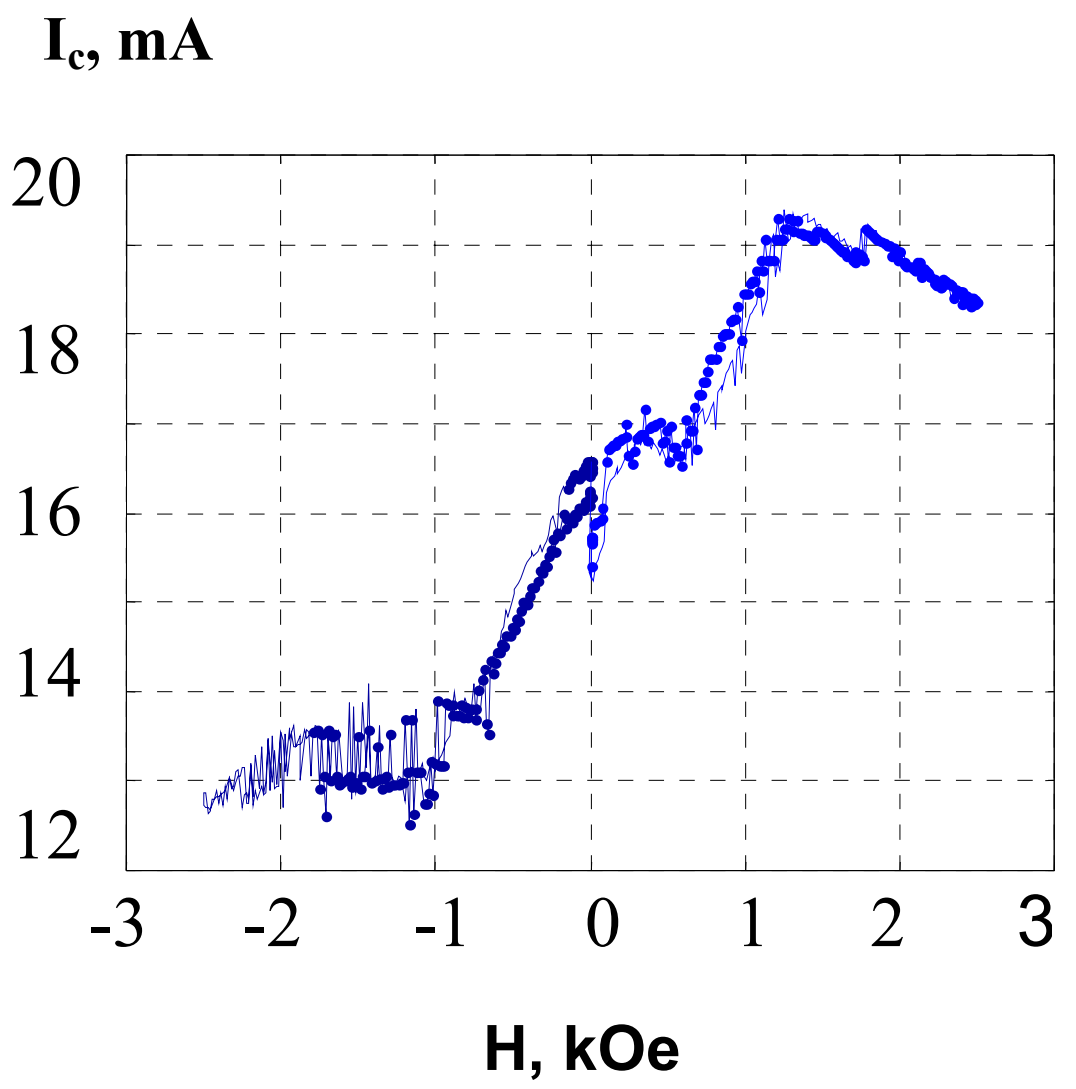

Fig. 6. Dependence $I_c(H)$ of the microbridge with ferromagnetic nanoparticles.

Today we do not have a clear understanding of the mechanism behind the magnetic particles effect on the critical current of a microbridge, all existing hypotheses have some imperfections. It seems important, though, that our *Nb* film is polycrystalline and features a high density of the critical current, and the microbridge of our make is narrow, which determines the boundary condition for entrance of the vortices.

## CONCLUSION

We investigated novel types of hybrid structures composed of ferromagnetic nanoparticles on superconductor, which can be used without control of the temperature. The transport properties of superconductors are well controlled by the magnetic state of the particles. For the case of the edge-type Josephson junction a simple model of the influence produced by magnetic particles is proposed. The experiments confirmed susceptibility of the Josephson junctions to the magnetic state of nanoparticles, which can be used in low temperature electronic devices.

The experimental results obtained in the study of overlap geometry junction and narrow microbridge are novel, and there is no theory interpreting these effects nowadays. However, the observed effects are interesting both from the fundamental point of view and for prospective application.

This work was supported by the RFBR, Grant number 03-02-16774, 04-02-16827, 04-02-17048 and INTAS, Grant number 03-51-6426, 03-51-4778.